# Solar Control on Jupiter's Equatorial X-ray Emissions: 26-29 November 2003 XMM-Newton Observation

Anil Bhardwaj[1,*], Graziella Branduardi-Raymont[2], Ronald F. Elsner[1], G. Randall Gladstone[3], Gavin Ramsay[2], Pedro Rodriguez[4], Roberto Soria[2], J. Hunter Waite Jr.[5], and Thomas E. Cravens[6]

[1] NASA Marshall Space Flight Center, NSSTC/XD12, 320 Sparkman Drive, Huntsville, AL 35805; anil.bhardwaj@msfc.nasa.gov
[2] Mullard Space Science Laboratory, University College London, Holmbury St Mary, Dorking, Surrey RH5 6NT, UK
[3] Southwest Research Institute, San Antonio, TX 78228
[4] XMM-Newton SOC, Apartado 50727, Villafranca, 28080 Madrid, Spain
[5] Dept. of Atmospheric, Oceanic, & Space Sciences, University of Michigan, Ann Arbor, MI 48109
[6] Dept. of Physics & Astronomy, University of Kansas, Lawrence, KS 66045

[*] On leave from: Space Physics Laboratory, Vikram Sarabhai Space Centre, Trivandrum 695022, India; bhardwaj_spl@yahoo.com

## Abstract

During November 26-29, 2003 XMM-Newton observed soft (0.2-2 keV) X-ray emission from Jupiter for 69 hours. The low-latitude X-ray disk emission of Jupiter is observed to be almost uniform in intensity with brightness that is consistent with a solar-photon driven process. The simultaneous lightcurves of Jovian equatorial X-rays and solar X-rays (measured by the TIMED/SEE and GOES satellites) show similar day-to-day variability. A large solar X-ray flare occurring on the Jupiter-facing side of the Sun is found to have a corresponding feature in the Jovian X-rays. These results support the hypothesis that X-ray emission from Jovian low-latitudes are solar X-rays scattered from the planet's upper atmosphere, and suggest that the Sun directly controls the non-auroral X-rays from Jupiter's disk. Our study also suggests that Jovian equatorial X-rays can be used to monitor the solar X-ray flare activity on the hemisphere of the Sun that is invisible to space weather satellites.

## 1. Introduction

X-ray emission from Jupiter is the brightest among planetary bodies in the solar system [*Bhardwaj et al*., 2002]. Jovian X-rays are basically of two types: 1) the "auroral" emissions, which are confined to high-latitudes (~>60º) in both polar regions, and 2) the "dayglow" emissions, which are from the low-latitude (~<50º) regions of the disk. (Henceforth, we define X-rays from the low-latitude regions of Jupiter as "disk" emissions.) With the recent X-ray observations of Jupiter by Chandra and XMM-Newton, it is now believed, as proposed earlier [see *Bhardwaj and Gladstone*, 2000 for review], that auroral X-rays of Jupiter are produced by charge-exchange of precipitating highly-ionized heavy ions. However, these ions are not from the inner (~8–12 $R_J$; $R_J$ = Jupiter radius) magnetosphere, as contemplated earlier, but are from the outer (>30 $R_J$) magnetosphere or/and solar wind, and apparently undergo acceleration to attain energies of >1 MeV/nucleon before impacting on the Jovian upper atmosphere [cf. *Gladstone et al*., 2002; *Cravens et al*., 2003; *Bhardwaj*, 2003; *Elsner et al*., 2004; *Branduardi-Raymont et al*., 2004].

*Waite et al*. [1997] reported the detection of X-rays from Jupiter's equatorial latitudes using the ROSAT/HRI, and suggested that they may be largely due to the precipitation of energetic sulfur and oxygen ions into the atmosphere from the radiation belts. However, *Maurellis et al*. [2000] showed that elastic scattering of solar X-rays by atmospheric neutrals and fluorescent production of carbon K-shell X-rays from methane molecules located below the Jovian homopause are also potential sources. Their model brightness agrees within a factor of two with the bulk low-latitude ROSAT/PSPC measurements, implying that solar photon scattering (~90% elastic scattering) may act in conjunction with energetic heavy ion precipitation to generate Jovian non-auroral X-ray emission. The solar X-ray scattering mechanism is also indicated from the correlation of Jovian X-ray emissions measured by ROSAT/HRI with the F10.7 cm solar flux [*Gladstone et al*., 1998]. Recently, Chandra and XMM-Newton have obtained spectra of the Jovian disk X-rays, which show that the disk emission is harder and extends to higher energies than the auroral spectrum [*Bhardwaj et al*., 2004; *Branduardi-Raymont et al*., 2004], suggesting that the processes producing X-rays are different at these two latitudes on Jupiter. In this letter we present the first direct evidence that the Sun controls the X-rays from Jupiter's disk.

## 2. Observation Details

XMM-Newton [*Jansen et al*., 2001] observed Jupiter for a total of ≈250 ks between Nov. 25, 23:00 and Nov. 29, 12:00, 2003. The data presented here were acquired with the EPIC MOS-1 and -2 [*Turner et al*., 2001] and EPIC-pn [*Struder et al*., 2001] cameras, operated in Full Frame and Large Window mode, respectively. The thick filter was used with all three cameras to minimize the risk of optical contamination. Six pointing trims were carried out during the observation, which spanned two XMM-Newton revolutions (rev. 0726, 33 hr observing time; rev. 0727, 36 hr), separated by a 17 hr gap due to perigee passage. The data were analyzed with the XMM-Newton Science Analysis Software (SAS) v.6 using the standard procedures [cf. *Branduardi-Raymont et al*., 2004]. Photons collected during the sequence of pointings were re-registered in the frame of the planet's center before further analysis. Periods of enhanced particle background were identified by examining the time series of events with energy >10 keV from the cameras whole field of view (30' diameter). Exclusion of these high background periods leaves 210 ks of good quality data on which the rest of the analysis was carried out. During the observation, Jupiter's angular diameter was 36″, and its heliocentric and geocentric distances were 5.4 and 5.5 AU, respectively. The phase angle (Sun-Jupiter-Earth angle) of the observations was 10.3°, and the solar elongation (Sun-Earth-Jupiter angle) was between 76.7° and 79.8° during the observation.

**3. X-Ray Image of Jupiter**

Figure 1(a) shows the distribution of X-ray brightness over Jupiter during the Nov. 26-29, 2003 XMM-Newton observation, obtained by combining EPIC MOS-1, -2 and -pn events in the energy range 0.2-2 keV, where essentially all the planet's X-ray emission is detected. This image is generated from ≈59 hours (6 Jupiter rotations) of observation time. The north and south auroral zones are brighter than the planetary disk emission, which is quite uniform, in agreement with the Chandra observations [*Gladstone et al*., 2002; *Elsner et al*., 2004; *Bhardwaj et al*., 2004]. The brightness of the disk emission in the 0.2-2 keV range is measured to be 0.08 R, which is in concurrence with that expected theoretically for disk X-rays produced by scattering of solar X-rays during moderate (F10.7 ~160) solar activity [*Cravens and Maurellis*, 2001]. The X-ray luminosity in the 0.2-2 keV band from the non-auroral disk of Jupiter is 0.4 GW.

The image in Fig. 1(b) is the same as in Fig. 1(a) but for energies in the 0.7-2 keV range, where the contribution from auroral emissions is largely reduced; the auroral emissions peak ~0.6 keV, while the disk spectrum has a broad peak in the 0.6-1 keV range and extends to relatively higher energies [*Branduardi-Raymont et al*., 2004; *Elsner et al*., 2004; *Bhardwaj et al*., 2004]. Both images in Fig. 1 show apparent limb darkening (more on the anti-sunward side). Simulations demonstrate that this pattern is a consequence of the larger Point Spread Function (PSF) of XMM-Newton (about 15" Half Energy Width, or HEW) compared to that of Chandra (~0.8" HEW), and is thus consistent with the uniform X-ray disk observed by Chandra [*Gladstone et al*., 2002; *Elsner et al*., 2004; *Bhardwaj et al*., 2004]. This statement is further asserted by Figure 2, where the results of radiative transfer simulations of the solar reflected photons across the equator of Jupiter are shown as a function of wavelength for the same phase angle of the XMM-Newton observation. These calculations were made using the radiative transfer code of *Gladstone* [1988], and include only Compton (incoherent) scattering and absorption cross sections for $H_2$; fluorescence (e.g. by carbon atoms in $CH_4$, whose contribution is <10% [cf. *Maurellis et al*., 2000]) is not considered. Figure 2 shows that the scattered solar photons will produce an appearance of uniform brightness across the equatorial disk for wavelengths >0.3-0.4 nm (<3-4 keV), except very close to the anti-sunward limb. This figure also indicates that only at higher (>10 keV) energies the Jovian X-ray disk will show an asymmetry in the planet's brightness associated with the relative position of the sub-solar point, producing a limb darkening effect. This is due to the increasing dominance of Compton scattering over absorption with increasing energy, resulting in near-conservative scattering at wavelengths <0.1 nm.

**4. EPIC Timing Analysis**

Because the XMM-Newton telescope PSF is wider than Chandra's, mixing of auroral and disk events takes place over a larger area; thus we extract Jupiter's disk events from a region confined to be close to the equator (a rectangular box of 28" x 13") where we can expect negligible contamination from the auroral spots [cf. *Branduardi-Raymont et al.*, 2004]. With six re-pointings (cf. Section 2) Jupiter was kept on the same CCD all the time. EPIC and RGS continued to collect data through the times of re-pointing and settling (a total of ~600 s at every re-pointing). The photon positioning could have an error of ~2" at worst for those 600 s. Considering that Jupiter's disk is 36″, any inaccuracy in the lightcurve will be negligible. Time series of the 0.2-2 keV disk events, constructed for the three EPIC cameras in 100 s bins, are combined and re-binned in 5 min intervals. No background subtraction is applied, because Jupiter is foreground to the diffuse X-ray background, and the residual particle and instrumental contribution at these energies is <5% of Jupiter's emission. These time series can be directly compared with those observed by TIMED/SEE for the solar X-ray flux in the 0.2-2.5 keV range, given that <1% of the Jupiter X-ray emission detected by EPIC falls between 2.0-2.5 keV, and that the temporal behavior of the solar flux in the 2.0-2.5 keV band is similar to that in the 0.2-2.0 keV band.

**5. Time Series of Jovian and Solar X-rays**

In Figure 3 we present the time series of Jovian disk X-rays observed by EPIC (pn+MOS; 0.2-2.0 keV, or 0.6-6 nm) along with the solar X-ray fluxes from the GOES 10 and 12 satellites in the 0.1-0.8 nm (1.6-12.4 keV) and 0.05-0.40 nm (3.1-24.8 keV) bands, from the SOHO/SEM in bands 0.1-50 nm (0.02-12.4 keV) and 26-34 nm (0.036-0.047 keV), and from the TIMED/SEE in 0.5-6.5 nm (0.2-2.5 keV) band. The solar fluxes are normalized value at 1 AU, and all the data set are 5-min binned data, except for the TIMED/SEE data which are 3-min observation-averaged fluxes obtained every orbit (~12 measurements per day). The Jovian X-ray data are plotted after correcting for the difference in light travel time (4948 s) from Sun to Earth and from Sun to Jupiter to Earth.

Overall, the Jupiter disk X-rays show day-to-day variability similar to that of solar X-rays (compare curves in panel (d) and those in other panels of Fig. 3): both show a general rise from 26 to 29 Nov. The average increase in Jovian X-ray 0.2-2 keV flux from Nov. 26 to that on Nov. 29 is about 40%, which is the same as that seen in the TIMED/SEE 0.2-2.5 keV flux, and is quite similar to that observed in the GOES X-ray 1.6-12.4 keV flux. For comparison, the corresponding increase in the auroral emissions is only about 10%. The SOHO/SEM 0.1-50 nm and 26-34 nm fluxes show an increase of ~5%; this is understandable because the solar flux at lower energy dominates the 0.1-50 nm flux.

Unfortunately, as our observations covered two consecutive XMM-Newton revolutions, the largest solar flare during the Nov. 26-29 period, a C9.6-class flare around 0820 UT on Nov. 27, was missed during perigee passage. This flare occurred from sunspot 10508 (NOAA/USAF region) which was the most active region during the 24-30 Nov. 2003 period, producing 20 C-class flares over this time. The location of this sunspot during Nov. 26-29 was around 19°S latitude (the angular distance from the solar equator, measured north (N) or south (S) along the meridian) and drifted from ~15°W to 50°W central meridian distance (CMD, the angular distance in solar longitude measured from the central meridian). Since Jupiter was leading Earth and the Sun-Earth-Jupiter angle was 78º while the phase angle was 10º, sunspot region 10508 was constantly visible to both Earth and Jupiter.

Figure 3 suggests that the day-to-day variability in Jupiter's disk X-ray flux is at the same level as that in the solar X-ray flux when the two are compared in the same energy range. To explore whether short-term fluctuations in solar X-rays are associated with a similar trend in the Jovian disk X-rays, in Figure 4 we plot the 30-min binned lightcurves of Jovian X-rays and GOES 0.1-0.8 nm solar X-ray flux, as well as the lightcurve for the background X-rays. The background lightcurve is generated by extracting the events, in the 0.2-2 keV band, selected with the same extraction box as for the equatorial region, but outside the planet's disk, in a nearby region of sky empty of X-ray sources. The background lightcurve starts later than that of the Jupiter disk because for MOS1, which displays the lowest background, the first event occurs much later than for MOS2 and pn, thus the combined lightcurve can only start at ~35 ks into the plot.

The sunspot 10508 produced the next largest X-ray flare (peak time, duration) of class C3.6 (0913UT, 24 min) on Nov. 28, and clearly a peak showed up in the Jovian X-rays at the same time (location marked by black arrow in Fig. 4). This flare also caused a sudden ionospheric disturbance on Earth. It is interesting to note that not only the time of the peak of the flare but also the relative increase in the Jovian X-rays (~50% above the mean flux on Nov. 28) is consistent with the increase in the solar flux (a factor of ~2 above the mean flux on Nov. 28), when allowance is made that the solar flux at higher energies fluctuate more than that at lower energies. There are other smaller flares of shorter duration from this and other sunspots, which have corresponding spikes in the Jovian X-rays at the same time, but they are not that statistically significant.

There are large peaks in the time series of Jovian X-rays at epochs (e.g., marked by green arrows) with no corresponding peaks in the solar X-rays. The likely reason is that, due to the viewing geometry in Nov. 2003, Jupiter is able to see about 50% of the solar hemisphere (western side) not visible from Earth. The Jovian disk X-rays will respond, producing spikes/peaks in the XMM-Newton lightcurves, to solar flares occurring on that hemisphere of the Sun which is hidden from Earth's view. This could be a boon to space sciences, since observations of Jupiter disk X-rays could provide a monitor of flaring on the hemisphere of the Sun that is not observable from Earth space weather satellites. In view of this logic, we suggest that solar flares, unseen at Earth, occurred at the times marked by green arrows in Fig. 4 on the western (Earth-hidden) side of the Sun. These flares are likely to have been produced by sunspot region 10501, which was the most active region during the Nov. 17-23 period, when it produced 9 M-class and 7 C-class flares; and this sunspot was beyond the Earth-visible western solar terminator during Nov. 26-29, 2003.

## 6. Discussion and Conclusion

The major results presented above can be summarized as: 1) the day-to-day variability in Jupiter's disk X-ray flux is on the same scale of that in the solar X-ray flux, 2) the biggest solar X-ray flare on the Jupiter-facing side of the Sun has a matching feature in Jupiter's disk X-ray lightcurve, 3) the EPIC soft (0.2-2.0 keV) X-ray image of Jupiter shows a uniform intensity disk that is consistent with that expected for scattered solar X-rays and is in agreement with Chandra observations, and 4) the EPIC-measured X-ray brightness over the low-latitude disk is 0.08 R, which is in agreement with the model calculations based on scattering of solar X-rays from Jupiter's upper atmosphere.

Jovian auroral X-rays observed by Chandra during Dec. 18, 2000 and Feb. 24-26, 2003 show quasi-periodic oscillations on a time scale of ~40±20-min [*Gladstone et al.*, 2002; *Elsner et al.*, 2004]. We searched for signatures of pulsation in the Nov. 2003 disk X-rays by constructing the power spectral density (PSD) distribution. No periods are evident in the PSD, which is consistent with the expectation that Jovian disk X-ray emissions are solar scattered X-rays [*Branduardi-Raymont et al.*, 2004; *Elsner et al.*, 2004; *Bhardwaj et al.*, 2004]. Further support for the solar scattering mechanism comes from the lack of association of the disk X-ray flux, observed by Chandra, with the system III longitude and the surface magnetic field strength [*Bhardwaj et al.*, 2004]; this argues against the *Waite et al.* [1997] suggestion of ring current precipitation as the source mechanism for Jovian disk X-rays.

The results presented in this paper suggest that Jovian disk X-rays are most plausibly resonant (~90%) scattered solar X-rays with minor (<10%) fluorescent contribution. However, not all the incident solar X-rays in the 0.2-2.0 keV are scattered back. The geometric X-ray albedo of Jupiter over this energy range is ~$5 \times 10^{-4}$ [*Bhardwaj et al.*, 2004]. Thus, Jupiter's atmosphere acts like a hazy-diffuse mirror for the incident solar X-rays – scattering back roughly one in few thousand photons – enabling Jupiter to shine in soft X-rays. It is worth noting, though, that the albedo of Jupiter is not a constant function of energy (cf. Fig. 2), but it increases with increasing energy of the incident photons.

In summary, the study presented in this letter has revealed that Jovian X-ray disk emissions are regulated by processes occurring on the Sun. We suggest that the study of Jovian disk X-rays, during certain Jupiter phases, can reveal information on solar flaring from the regions/hemisphere of the Sun that are not visible from the Earth.

**Acknowledgements**

This work is based on observations obtained with XMM-Newton, an ESA science mission with instruments and contributions directly funded by ESA Member States and the USA (NASA). This research was performed while A. Bhardwaj held a National Research Council Senior Resident Research Associateship Award at NASA Marshall Space Flight Center. The MSSL authors acknowledge the financial support of the UK Particle Physics and Astronomy Research Council. We thank Tom Wood for help in providing the TIMED/SEE Version 7 Data Products. The solar SEM data are from the CELIAS/SEM experiment on the SOHO spacecraft which is a joint ESA and NASA mission. GOES X-ray data are obtained from Space Physics Interactive Data Resource site http://spidr.ngdc.noaa.gov/spidr/index.html.

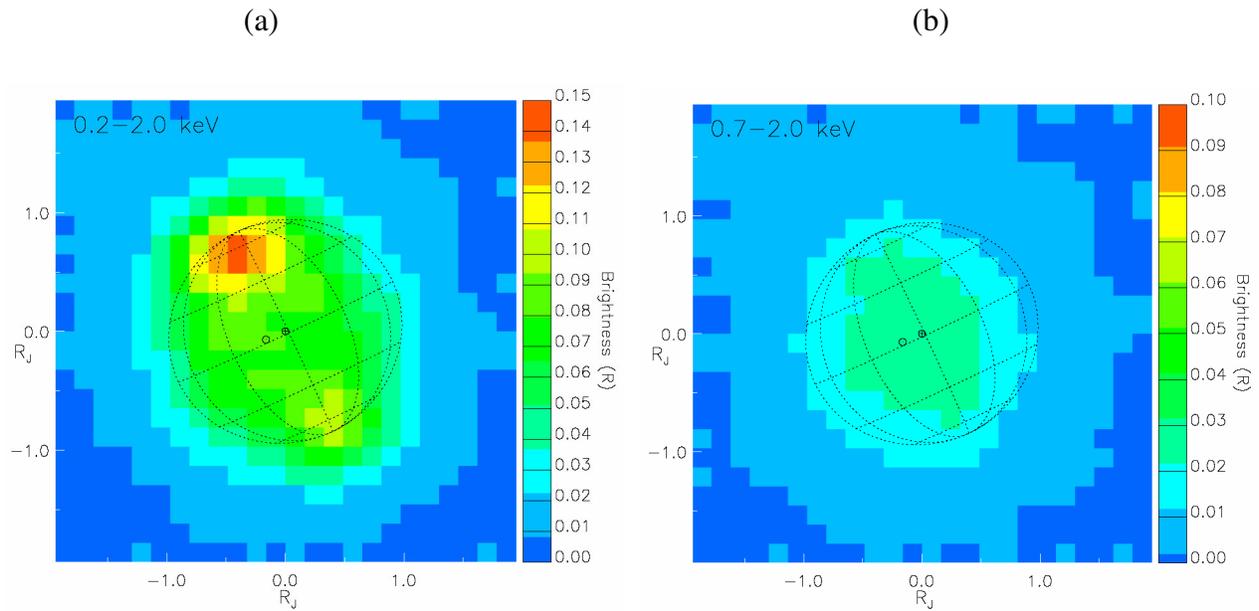

**Figure 1.** X-ray image of Jupiter during Nov. 26-29, 2003 observed by XMM-Newton, obtained by combining the events from EPIC-pn and both EPIC-MOS cameras. **(a)** Left image: 0.2-2.0 keV; **(b)** right image: 0.7-2 keV. The superimposed graticle shows latitude and longitude lines at intervals of 30°. The small circles near the center represent the sub-Earth and sub-solar points. The axes are labeled in units of Jupiter's radius, $R_J$, and the scale bar is in Rayleighs (R).

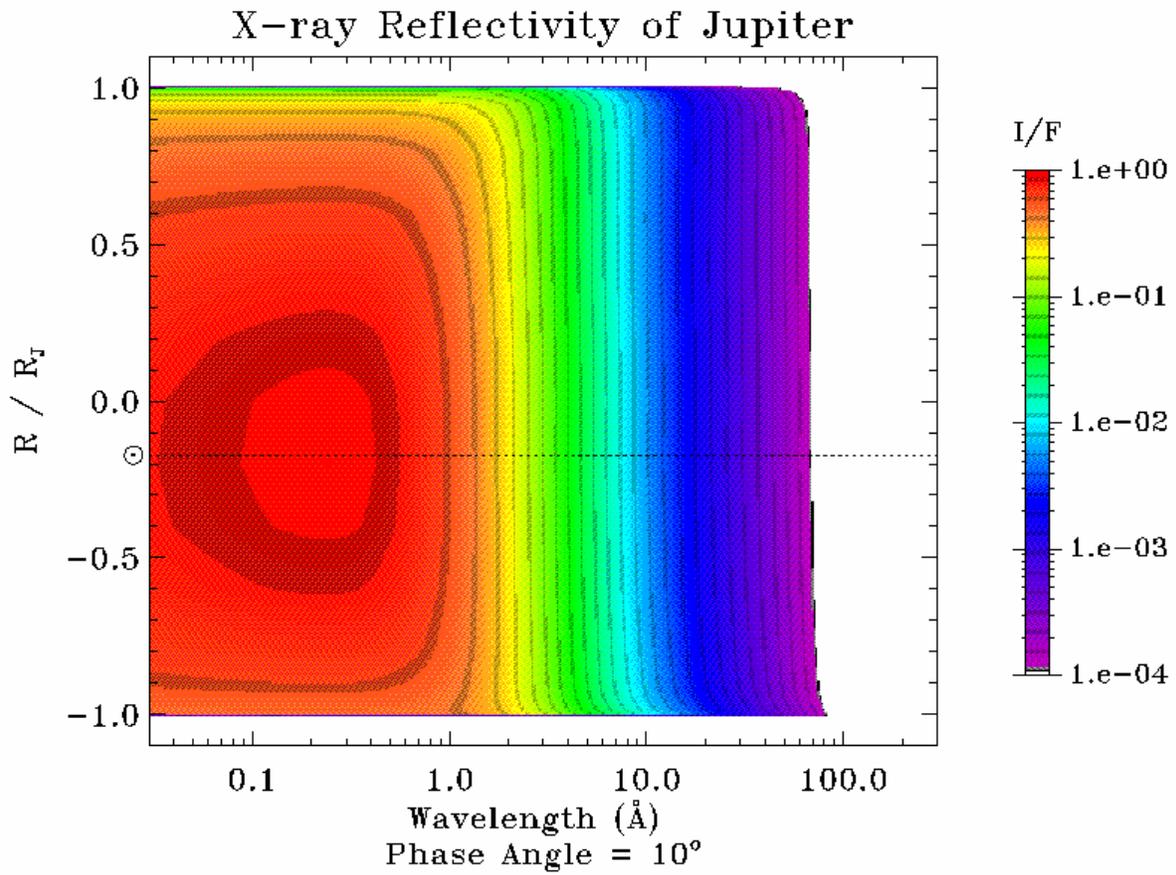

**Figure 2.** Model reflectivity across the Jovian equator as a function of wavelength, for a phase angle of 10°. The y-axis is in units of Jupiter's radius, $R_J$, and the log scale bar for I/F indicates the large increase in atmospheric reflectivity at shorter (<0.1 nm) wavelengths due to Compton scattering.

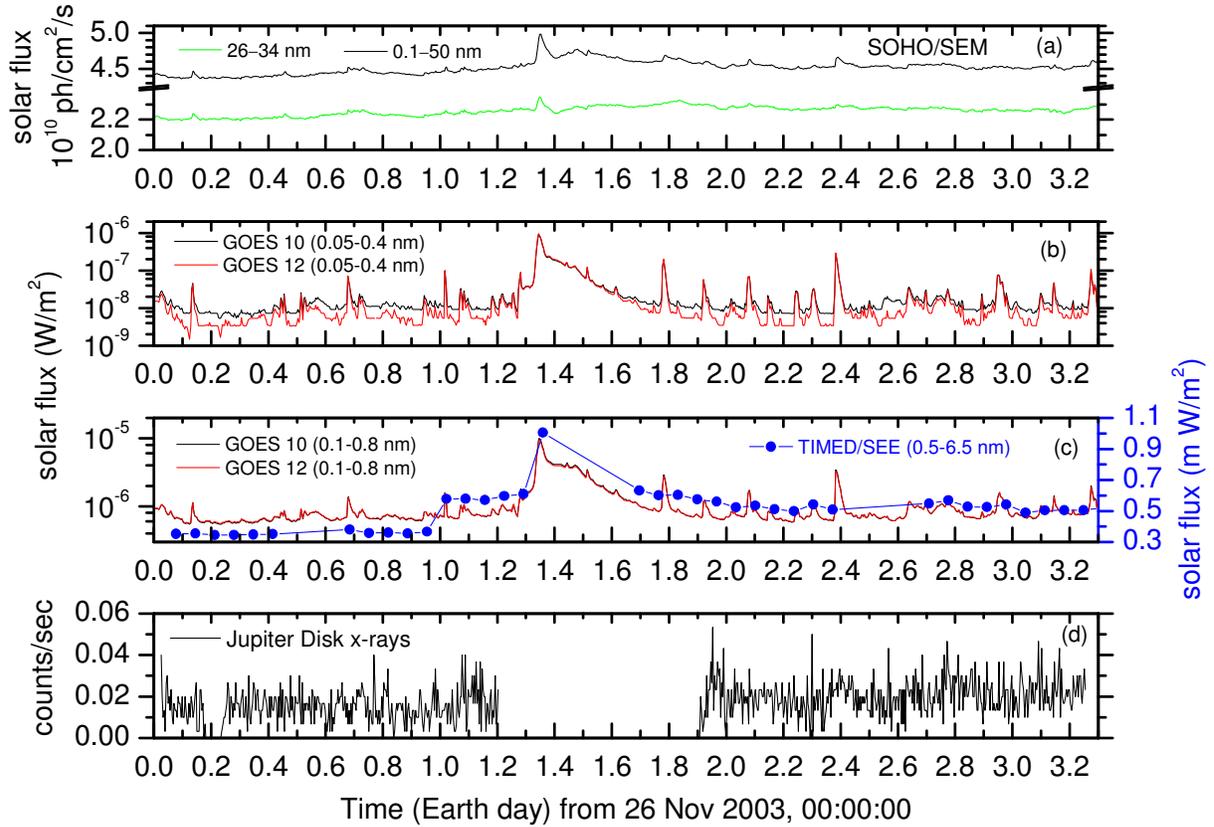

**Figure 3.** Time evolution of Jovian disk X-rays (0.6-6 nm) compared with solar X-ray-EUV flux during Nov. 26-29, 2003. All data are 5-min binned (except TIMED/SEE, see below) and solar fluxes are scaled values at 1 AU. **(a)** Solar 0.1-50 and 26-34 nm flux measured by SOHO/SEM. Note the break in the y-axis. **(b)** Solar 0.05-0.4 nm flux measured by GOES 10 and 12. **(c)** Solar 0.1-0.8 nm flux measured by GOES 10 and 12 (the black lightcurve showing GOES 10 data is overlapped by GOES 12 data, which is shown in red), and solar 0.5-6.5 nm 3-min observation-averaged flux, obtained every orbit, measured by TIMED/SEE. Note that TIMED/SEE data points (~12 per day) are filled blue circles that are connected by solid line for visualization purpose. **(d)** X-ray flux from Jupiter's low-latitude disk observed by XMM-Newton, plotted after shifting by -4948 s to account for light travel time delay between Sun-Jupiter-Earth and Sun-Earth. The small gap at ~0.2 days is due to a loss of telemetry from XMM-Newton, and the gap between ~1.2-1.9 days is caused by the satellite perigee passage.

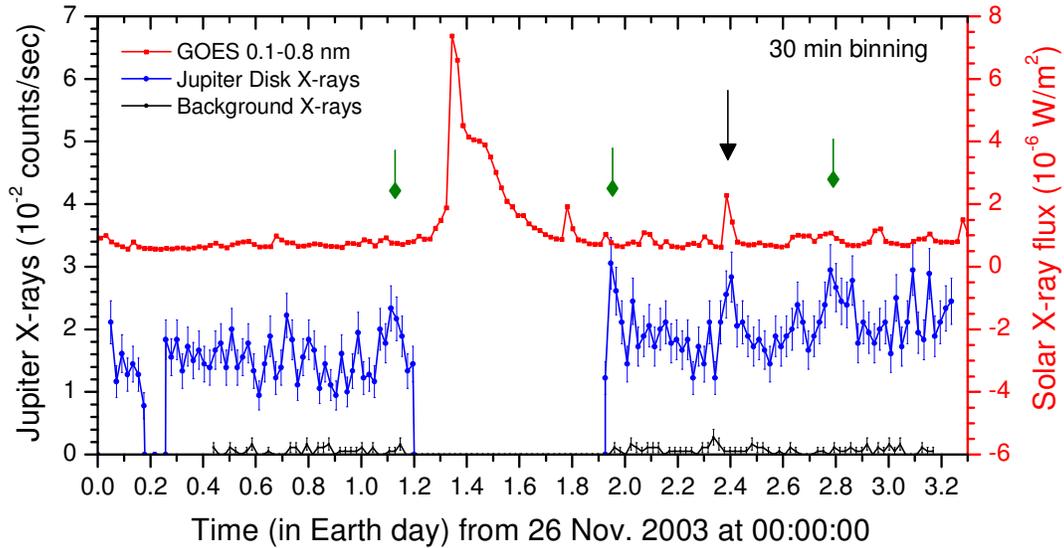

**Figure 4.** Comparison of 30-min binned Jupiter disk X-rays (solid blue circles, with error bars, joined by blue line) with GOES 10 0.1-0.8 nm (solid red squares joined by red line) solar X-ray data. The lightcurve of background X-rays is also shown (black line with error bars, at the bottom of the figure). The Jovian X-ray time is shifted by -4948 s to account for light travel time delay between Sun-Jupiter-Earth and Sun-Earth. The small gap at ~0.2 days is due to a loss of telemetry from XMM-Newton, and the gap between ~1.2-1.9 days is caused by the satellite perigee passage. The black arrow (at 2.4 days) refers to the time of the largest solar flare visible from both, Earth and Jupiter, during the XMM-Newton observation, which has a clear matching peak in the Jovian lightcurve. The green arrows represent times when the Jupiter lightcurve shows peaks, which we suggest correspond to solar flares that occurred on the western (Earth-hidden) side of the Sun.